\documentclass[preprint,12pt]{elsarticle}

\usepackage{amssymb}

\usepackage[dvipsnames]{xcolor}
\usepackage{color}
\usepackage[version=3]{mhchem} 
\usepackage[a-1b]{pdfx}
\usepackage{times}
\usepackage{amsmath}
\usepackage{amssymb}
\usepackage{bm}
\usepackage{here}
\usepackage{graphicx}
\usepackage{framed}
\definecolor{shadecolor}{rgb}{0.9,0.9,0.9}
\usepackage[utf8x]{inputenc}
\biboptions{sort&compress}

\journal{Mater. Today Phys.}

\begin{document}

\begin{frontmatter}



\title{Theoretical Predictions of MB$_5$N$_5$: Atom-Stuffed Boronitride Clathrate Cages
Derived from the High-Pressure Superhydride}


\author[label1]{Nisha Geng}

\affiliation[label1]{organization={Department of Chemistry},
            addressline={ State University of New York at Buffalo}, 
            city={Buffalo},
            postcode={14226-3000}, 
            state={NY},
            country={USA}}

\author[label1]{ Giacomo Scilla}

\author[label1]{Eva Zurek\corref{mycorrespondingauthor}}\cortext[mycorrespondingauthor]{Corresponding author}
\ead{ezurek@buffalo.edu}

\begin{abstract}
This study investigates 198 $MX_5Y_5$ ($X$, $Y$ = B, C, or N) clathrate-like structures derived from MH$_{10}$ superhydrides using high-throughput Density Functional Theory (DFT) geometry optimizations and phonon calculations. A wide variety of electropositive and electronegative encapsulated atoms were considered. From all of the studied systems only 34 $M$B$_5$N$_5$ phases were found to be dynamically stable at ambient pressure. The highest 1-atmosphere superconducting critical transition temperature was predicted for FB$_5$N$_5$. However, \textit{ab initio} molecular dynamics simulations revealed that all of the identified superconducting phases decompose by 300~K at 1~atm, while only eleven semiconducting phases remained thermally stable. Our findings underscore the critical role of kinetic and thermal stability in predicting viable superconductors. The electronic structure of the $M$B$_5$N$_5$ compounds were rationalized in terms of electron donating and withdrawing intercalants. DFT and machine-learning based predictions of their mechanical properties were compared with those of an empty boronitride cage. 
\end{abstract}



\begin{keyword}
First-Principles Calculations  \sep Boronitrides \sep Superconductivity \sep Mechanical Properties \sep Electronic Structure



\end{keyword}

\end{frontmatter}


\newpage

\section{Introduction}

Building on Neil Ashcroft's ``chemical precompression" prediction that doping hydrogen with a second element could lower the pressure required for its metallization and induce superconductivity, significant progress has been achieved in the search for hydrogen-rich materials exhibiting a high superconducting critical temperature ($T_c$). Several hydride phases with high $T_c$s have been successfully synthesized under pressure using diamond anvil cells, including H$_3$S ($T_c=$~203~K at 150~GPa~\cite{Drozdov:2015}), various $M$H$_6$ compounds where $M$ is an electropositive element ($e.g.$\ CaH$_6$ [$T_c=$~215~K at 172~GPa~\cite{Ma:CaH6}, or 210~K at 160~GPa~\cite{Li:CaH6}], (La,Y)H$_6$ [$T_c=$~237~K at 183~GPa~\cite{Semenok:2021}], and YH$_6$ [$T_c=$~220~K at 183~GPa~\cite{Kong:2021a}, or 224~K at 166~GPa~\cite{Troyan-YH4}), $M$H$_9$ ($e.g.$ YH$_9$ [$T_c=$~243~K at 201~GPa~\cite{Kong:2021a}), and, relevant for our study, $M$H$_{10}$ superhydrides ($e.g.$ LaH$_{10}$ [$T_c=$~260~K at 200~GPa~\cite{Somayazulu:2019,Drozdov:2019}]). Many of these superconductors were initially proposed theoretically, underscoring the critical role of first-principles-based methods such as crystal structure prediction (CSP) searches and electron-phonon coupling (EPC) calculations in the discovery of new superconducting candidates\cite{Zurek:2018m,Zurek:2020k,Flores-Livas:2020}.

Although achieving room-temperature superconductivity is no longer a dream with the discovery of these hydrogen-based superconductors, a major obstacle still remains: none of the predicted or synthesized high-$T_c$ superhydrides found at high pressure are stable or recoverable at ambient pressure -- they decompose into classic metal hydrides and molecular H$_2$. One promising direction is to find light-element-based compounds whose structures are analogous to the high-pressure hydride superconductors, but whose strong covalent bonds are kinetically protected from overcoming barriers that would lead to their decomposition at ambient pressures. Conceptually, these systems can be constructed by replacing either some or all of the weakly covalently bonded hydrogen framework atoms in $M$H$_6$, $M$H$_9$, and $M$H$_{10}$ clathrates with light elements such as boron, carbon, nitrogen and even silicon, with lighter elements typically being associated with higher $T_c$s.

One strategy focuses on replacing  only part of the hydrogen framework. For instance, by substituting the hydrogen atoms at the 8$c$ site with boron, carbon, or nitrogen atoms in the $Fm\bar{3}m$ $M$H$_{10}$ structure, several $Fm\bar{3}m$ $M$X$_2$H$_8$ compounds~\cite{Gao:2021, Durajski:2021,Zurek:2022f,Wan:2022, Geng:2024s, Li:2022s, Jiang:2022} were predicted computationally to persist to lower pressures, including KB$_2$H$_8$ ($T_c=$~146~K at 12~GPa)~\cite{Gao:2021}, YC$_2$H$_8$ ($T_c=$~61~K at 50~GPa)~\cite{Zurek:2022f}, and AlN$_2$H$_8$ ($T_c=$~118~K at 40~GPa)~\cite{Wan:2022}. Alternatively, several $M$XH$_8$ phases were proposed by removing a hydrogen atom from the 8$c$ site of $M$H$_{10}$ and then replacing one of the hydrogen atoms by Be or another light $p$-block element~\cite{Zhang:2021,Lucrezi:2022,Liang:2021a,Di:2021bh}. Promisingly, one compound from this family, LaBeH$_8$, was recently synthesized with a measured $T_c$ of 110~K at 80~GPa~\cite{Song:2023s}. A second approach is to replace the whole hydrogen framework by light elements. For instance, $Pm\bar{3}n$ SrB$_3$C$_3$, a borocarbide analogue of the $M$H$_6$ clathrate, was found via CSP techniques~\cite{Li:2020sr} and predicted to be thermodynamically stable between 50-200~GPa \cite{Li:2020sr} with an estimated $T_c$ of 43~K at ambient pressure~\cite{zhu:arxiv}. This bipartite sodalite (Type-VII clathrate) compound was synthesized at 57~GPa, and recovered to ambient conditions with a measured onset $T_c$ of $\sim$20~K at~40~GPa~\cite{Li:2020sr,zhu:arxiv}. Isotypic LaB$_3$C$_3$ was also  synthesized at milder pressures and quenched to ambient conditions, but it was computed to be a semiconductor with a bandgap of 1.3~eV with the HSE06 hybrid functional~\cite{Strobel:2021la}. Since then, a number of borocarbide clathrates with two different metal atoms have been studied theoretically~\cite{Di:2022,Zhang:2022,Gai:2022, Geng:2023}; from these KPbB$_6$C$_6$ was predicted to have the highest $T_c$:  88~K~\cite{Geng:2023} at ambient pressure. Analogous sodalite-like boronitrides have also been extensively studied by theory~\cite{Li:2019, Hai:2022i}, and an AlB$_3$N$_3$ phase was predicted to have a $T_c$ of 72~K at ambient pressure, while borosilicides were computed to have lower $T_c$s owing to the heavier mass of the $p$-block element ($e.g.$\ RbB$_3$Si$_3$ with $T_c=$14~K)~\cite{Cui:2020}. Finally, boron-carbon clathrate cages that are structurally related to the Type-I and Type-II clathrates have been proposed~\cite{Bi:2024}.
\begin{figure*}
\begin{center}
\includegraphics[width=0.9\columnwidth]{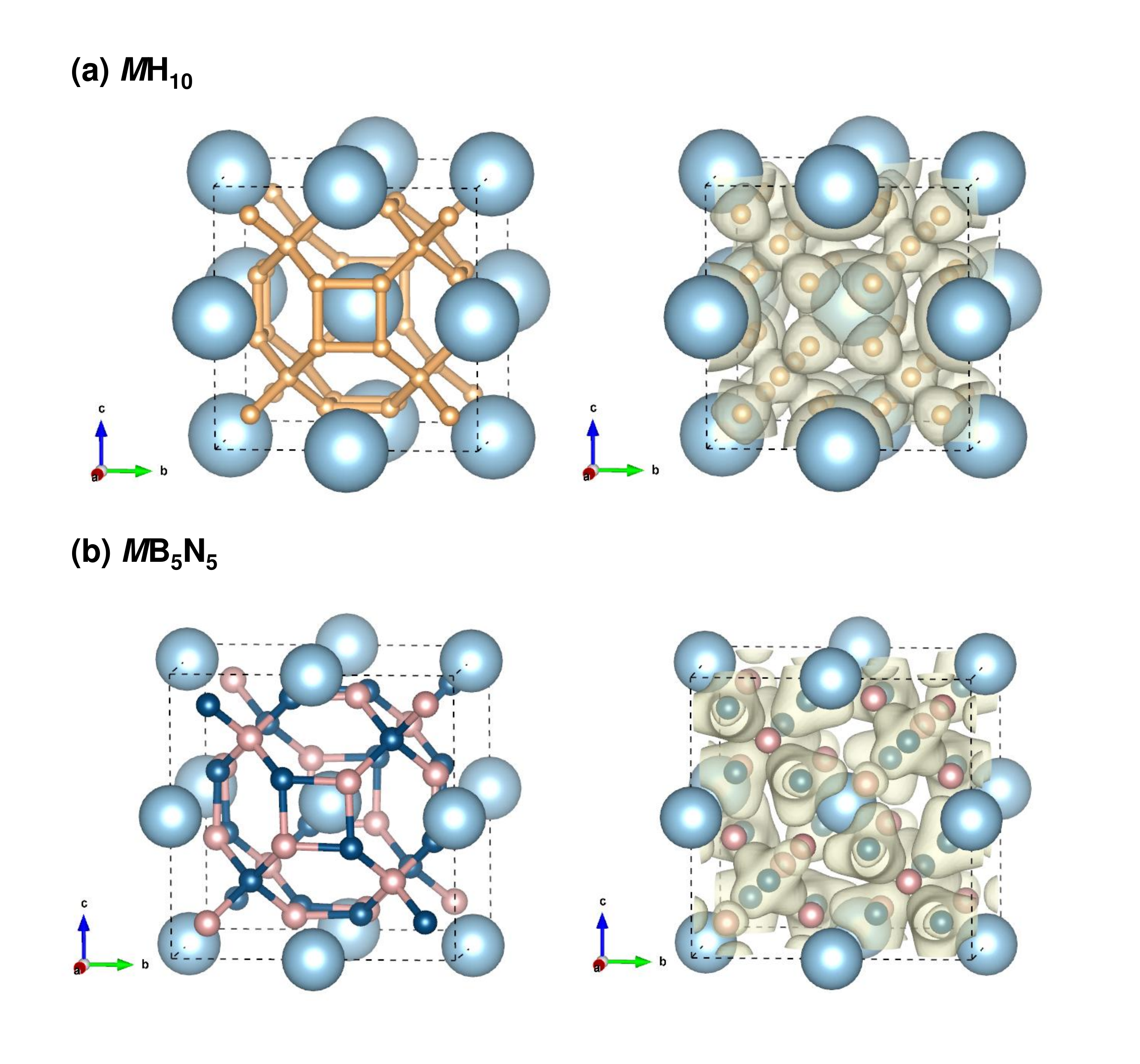}
\caption{Illustrations of the structure (left) and electron localization function (ELF) plot (right) of the conventional cell of: (a) $Fm\bar{3}m$ $M$H$_{10}$ and (b) $F\bar{4}3m$ $M$B$_5$N$_5$. $M$/H/B/N atoms are colored teal/orange/pinkish/navy blue. An isovalue of 0.55 and 0.7 was employed for generating the ELF plot computed for $Fm\bar{3}m$ LaH$_{10}$ at 300~GPa and $Fm\bar{3}m$ NaB$_5$N$_5$ at 1 atmosphere (atm), respectively.}
    \label{fig:mx10structure}
\end{center}
\end{figure*}

While high-throughput density functional theory (DFT) calculations have considered a majority of the possible substitutions of \emph{all} of the hydrogen atoms by $p$-block elements in $M$H$_6$ stoichiometry systems, most of the published studies have only considered \emph{partial} substitutions of the hydrogen atoms in the $Fm\bar{3}m$ $M$H$_{10}$ framework (Figure\ \ref{fig:mx10structure}(a)). To date, only a few investigations have examined replacing all of the hydrogen atoms via boron, carbon or nitrogen in $M$H$_{10}$. Similar to $M$X$_2$H$_8$ compounds~\cite{Gao:2021, Durajski:2021,Zurek:2022f,Wan:2022, Geng:2024s, Li:2022s, Jiang:2022}, Li \emph{et al.}\ replaced the hydrogen atoms at the $8c$ site with boron and those at the $32f$ site with carbon, designing a series of $Fm\bar{3}m$ $M$B$_2$C$_8$ compounds, with TlB$_2$C$_8$ exhibiting the highest predicted $T_c$ of $\sim$96~K at ambient pressure~\cite{Li:2024h}. Furthermore, Ding \emph{et al.}\ proposed two boronitride $M$H$_{10}$-like structures by occupying the $8c$ and $32f$ sites with equal numbers of boron and nitrogen atoms as shown in Figure\ \ref{fig:mx10structure}(b)~\cite{Ding:2022a}. The resulting $F\bar{4}3m$ YB$_5$N$_5$ and LaB$_5$N$_5$ compounds were computed to have a $T_c$ of 69~K and 59~K, respectively, at ambient pressure. However, the rest of the map of possible $M$ atom substitutions in borocarbide and boronitride cages derived from the $Fm\bar{3}m$ $M$H$_{10}$ clathrates remains unexplored. 

Herein, we performed a high-throughput DFT screening where boron, carbon, or nitrogen atoms were employed to construct the clathrate framework yielding $F\bar{4}3m$ $MX_5Y_5$ ($X, Y$ = B, C, or N) stoichiometry compounds with a wide variety of enclathrated elements, $M$, at ambient pressure. We evaluated their dynamic stability, electronic structure and superconducting properties. Out of the 198 different combinations considered, 34 with $M$B$_5$N$_5$ stoichiometry were predicted to be dynamically stable (18 were computed to be metals and 16 semiconductors), whereas all of the $M$B$_5$C$_5$ or $M$C$_5$N$_5$ combinations were found to be dynamically unstable within the harmonic approximation. Among the metallic phases, group 16 elements possessed the highest (DOS) at the Fermi level ($E_F$), but small EPC constants ($\lambda  \le$~0.6) , resulting in predicted $T_c$ values below 10~K. The highest 1~atm $T_c$ was calculated for FB$_5$N$_5$ ($T_c=$~75~K). However, molecular dynamics simulations suggested that all of the superconducting $M$B$_5$N$_5$ phases would undergo structural transformations at 300~K. Only 11 of the semiconducting $M$B$_5$N$_5$ phases, including $M=$~He, Ne, Ar, Kr, Xe, Pd, Cu, Ag, Zn, Cd, and Hg, were found to be thermally stable, highlighting the importance of considering kinetic and thermal stability in future predictions of conventional superconductors.

\section{Computational Details}
Geometry optimizations, molecular dynamics simulations, and electronic structure calculations including band structures, densities of states (DOS), electron localization functions (ELF), and Bader charges were performed by density functional theory (DFT) as implemented in the Vienna \textit{ab-initio} Simulation Package (VASP) version 6.3 \cite{Kresse:1993a, Kresse:1999a}, with the gradient-corrected exchange and correlation functional of Perdew{-}Burke{-}Ernzerhof (PBE) \cite{Perdew:1996a}. The projector augmented wave (PAW) method \cite{Blochl:1994a} was employed, and the valence electron configurations listed in Table S1-2 were treated explicitly in all of the calculations. The plane-wave basis set energy cutoff was 900~eV and the B 2s$^2$2p$^1$, C 2s$^2$2p$^2$, and N 2s$^2$2p$^3$ electrons were treated explicitly in all of the non-spin-polarized calculations. The valence configurations and PAWs employed for the other elements are listed in Table S1-2. The $k$-point meshes were generated using the $\Gamma$-centered Monkhorst-Pack scheme and the number of divisions along each reciprocal lattice vector was selected so that the product of this number with the real lattice constant was 50~\AA{} in the geometry optimizations and 70~\AA{} otherwise. The optimized structural parameters are provided in Table S8. The crystal orbital Hamilton population (COHP) and the negative of the COHP integrated to the Fermi level (-iCOHP) was calculated using the LOBSTER package to analyze the bonding of selected phases~\cite{Maintz:2016}. The thermal stability of selected $M$B$_5$N$_5$ phases was examined by performing \emph{ab initio} molecular dynamics (AIMD) simulations using a canonical $NVT$ ensemble at 300~K, with temperature and volume controlled via a Nos\'e-Hoover thermostat~\cite{Nose:1984,Shuichi:1991,Hoover:1985,Frenkel:2023}. A $2\times2\times2$ supercell was chosen to reduce the constraint of periodic boundary conditions with an energy cutoff of 600~eV, and only the $\Gamma$-point was used. All AIMD simulations contained 10,000~MD steps (10~ps). High-throughput phonon calculations were carried out through a supercell approach~\cite{Parlinski:1997,Chaput:2011} using the VASP package coupled to the PHONOPY code~\cite{Togo:2015} with 2$\times$2$\times$2 supercells.

The electron-phonon coupling (EPC) calculations were performed with the Quantum Espresso (QE) program package~\cite{Giannozzi:2009}. The pseudopotentials employed, listed in Table S1, were obtained from the PSlibrary \cite{DalCorso:2014}, and generated by the method of Trouiller-Martins \cite{Troullier:1991} with the PBE functional~\cite{Perdew:1996a}. The Brillouin zone sampling $\Gamma$-centered Monkhorst-Pack~\cite{Monkhorst:1976} scheme was applied using Methfessel-Paxton~\cite{Methfessel:1989} smearing with a broadening smearing width of 0.006-0.01~Ry as provided in Table S3. A $16\times16\times16$ $k$-point grid was used for all phonon calculations, while a dense $32\times32\times32$ $k$-point grid and an $8\times8\times8$ $q$-mesh was used for all of the EPC calculations. The EPC parameter, $\lambda$, was calculated using a set of Gaussian broadenings from 0.0 to 0.500~Ry (with an increment of 0.005~Ry) and converged to 0.05~Ry. The $T_c$s were estimated using the Allen-Dynes modified McMillan equation \cite{Allen:1975} with a renormalized Coulomb potential, $\mu^*$, of 0.1, along with numerical solution of the Eliashberg equations~\cite{Eliashberg:1960} for selected systems. 

\section{Results and Discussion}

\begin{figure*}
\begin{center}
\includegraphics[width=1\columnwidth]{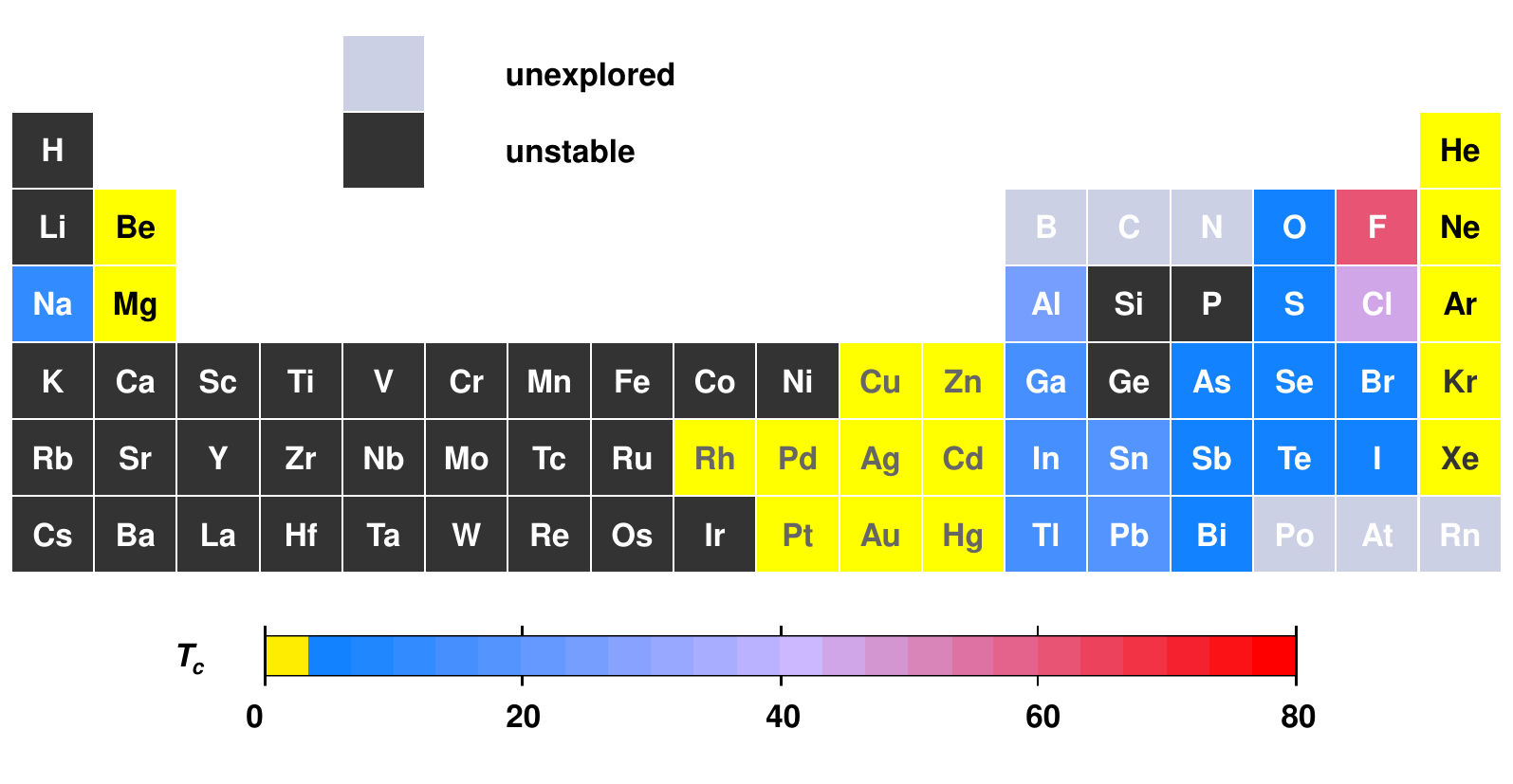}
\caption{A portion of the periodic table where the squares enclosing the elemental blocks are shaded according to the superconducting critical temperatures, $T_c$s, calculated via numerical solution of the Eliashberg equations and assuming a Coulomb repulsion parameter, $\mu^*=$~0.1, for $M$B$_5$N$_5$ compounds at ambient pressure. The identities of the center $M$ atoms are given within the squares. Phases that are dynamically unstable from harmonic phonon calculations are colored in black, unexplored phases are colored in grey, and insulating phases are colored yellow.}
    \label{fig:tc}
\end{center}
\end{figure*}

DFT calculations were performed to explore the boron, carbon, or nitrogen clathrates with $MX_5Y_5$ ($X, Y$ = B, C, or N) stoichiometries. The various enclathrated $M$ elements resulted in a wide range of valence electron concentrations that could potentially induce metallicity in the otherwise insulating boronitride cages. Phonon calculations were employed to assess their dynamic stabilities. Out of the 198 $MX_5Y_5$ (X, Y = B, C, or N) phases relaxed at ambient pressure, 34 of the $M$B$_5$N$_5$ combinations including $M$ = Na, Be, Mg, Al, Ga, In, Tl, Sn, Pb, As, Sb, Bi, O, S, Se, Te, F, Cl, Br, I, He, Ne, Ar, Kr, Xe, Rh, Pd, Pt, Cu, Ag, Au, Zn, Cd, and Hg were found to be dynamically stable, while all of the $M$B$_5$C$_5$ and $M$C$_5$N$_5$ phases were dynamically unstable (results are summarized in Figure S1). Contrary to the predictions of Ding \emph{et al.}, who concluded that YB$_5$N$_5$ and LaB$_5$N$_5$ were dynamically stable at ambient pressure with $T_c$s of 69 K and 59 K \cite{Ding:2022a}, respectively, our study showed that these phases do not correspond to local minima at these conditions (Figure S11). Our results for $M$B$_5$N$_5$ are graphically summarized in Figure\ \ref{fig:tc}, with dynamically stable phases shown as rectangles color-coded according to their predicted $T_c$ (from blue/cold to red/hot). Insulating compounds are colored yellow, and those that are dynamically unstable are colored black. Po, At and Rn atoms were not considered as they are radioactive and unlikely to be studied experimentally, and neither were B, C or N atoms as there have been numerous studies predicting B-C-N polymorphs~\cite{Zhang:2013a}. Furthermore, due to the computational expense involved potential magnetism was not considered in our high-throughput screenings. The subsequent discussion will focus exclusively on the 1~atm dynamically stable $M$B$_5$N$_5$ compounds. 

\begin{figure*}
\begin{center}
\includegraphics[width=0.85\columnwidth]{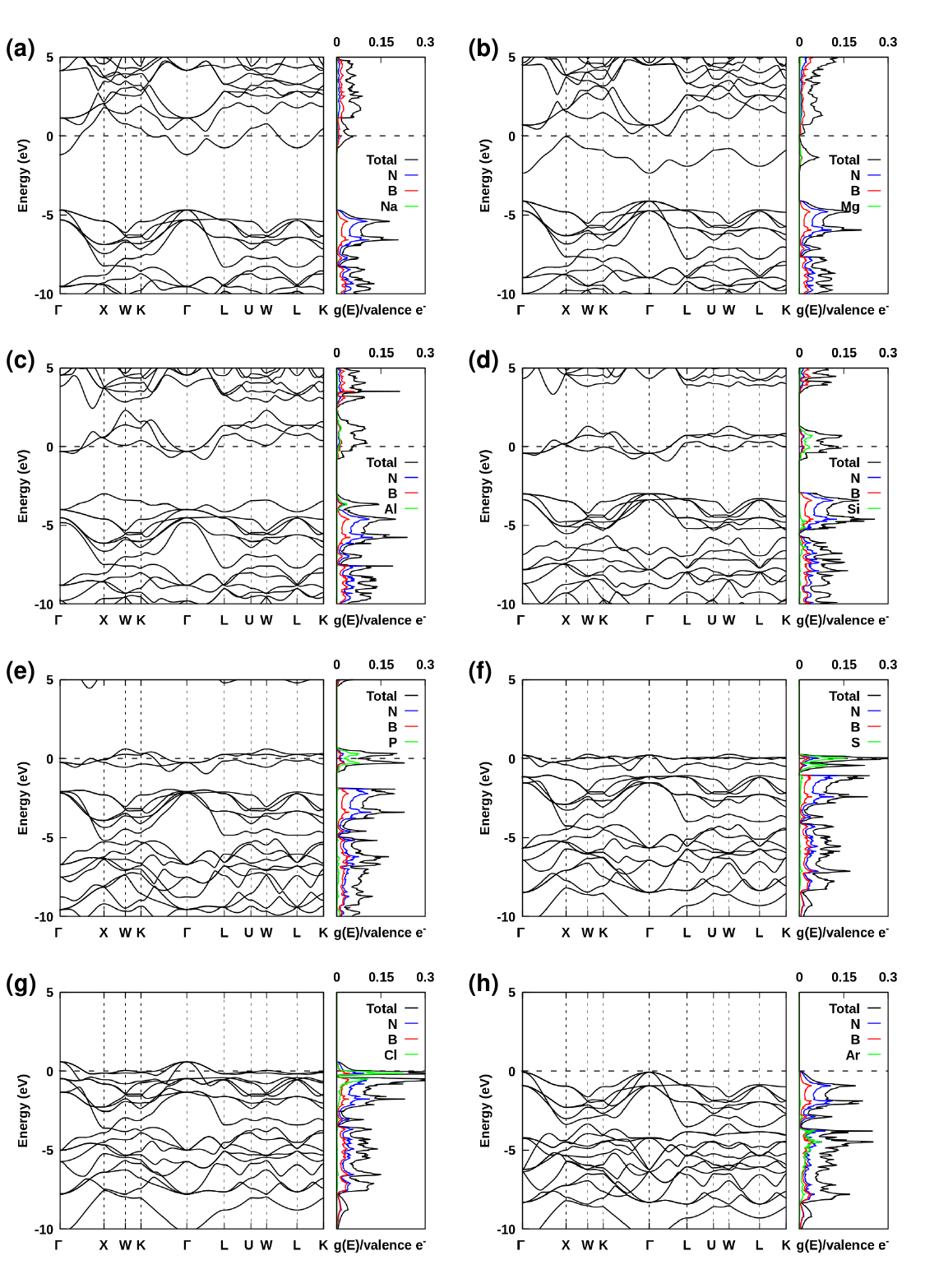}
\caption{Electronic band structures (at the PBE level of theory) and projected DOS for $M$B$_5$N$_5$ phases that contain intercalating elements, $M$, with varying numbers of valence electrons: (a) NaB$_5$N$_5$, (b) MgB$_5$N$_5$, (c) AlB$_5$N$_5$, (d) SiB$_5$N$_5$, (e) PB$_5$N$_5$, (f) SB$_5$N$_5$, (g) ClB$_5$N$_5$, and (h) ArB$_5$N$_5$. The top of the valence band (h) or Fermi energy (elsewhere) is set to 0~eV.}
    \label{fig:dos}
\end{center}
\end{figure*}

The empty B$_5$N$_5$ clathrate cages are isoelectronic with C$_{10}$, and within them each atom is 4-coordinate. As a result, one would expect that the vacant cages are wide gap insulators, as is diamond, which was verified by our calculations. Therefore, placing electropositive elements within the cages is expected to electron dope them, whereas electronegative elements such as fluorine or chlorine could, in principle, withdraw electrons from the boronitride cages, resulting in an enclathrated closed shell halide. The insertion of noble gases, on the other hand, is unlikely to result in any charge transfer, however it might serve to further localize electrons within the cage atoms, thereby opening the gap, as has been computed for a variety of other compounds upon uptake of helium (See Reference \cite{Zurek:2023b} and Refs.\ within). A further possibility is that encapsulation of the inert gases affects the mechanical properties as has been found in A-site vacant perovskites that are filled with helium~\cite{Zurek:2023b,Hester_2017_JACS,Lloyd_2021_ChemofMat}. These aspects will be explored in detail below.

The DOS at $E_F$ is often used as a descriptor for the $T_c$ of conventional superconductors, as it is related to the number of electrons available to participate in the superconducting mechanism~\cite{Shipley:2021a}. By varying the identity of the intercalated elements, one can identify the best candidates to achieve the highest DOS at $E_F$, which is assumed to be correlated with the highest $T_c$. This strategy has been applied in several high-throughput studies, such as those performed on $MX_2$H$_8$~\cite{Geng:2024s, Wan:2022}, $XY$B$_6$C$_6$\cite{Geng:2023, Di:2022, Zhang:2022} and $XY$B$_6$N$_6$~\cite{Hai:2022i} stoichiometry systems. To explore how the identity of the enclathrated elements modifies the DOS at $E_F$ we plotted the electronic bands and DOS of the studied systems; a selection of these for row 3 elements is provided in Figure\ \ref{fig:dos}. Enclathrated electropositive elements were computed to have positive Bader charges, such as +0.74, +0.84 and +0.69 for Na, Mg and Al (Table S5), suggesting that they donate electrons to the clathrate lattice, as expected.  For Na half a band (which would be empty in the B$_5$N$_5$ cage) is filled, and for Mg a full band is filled resulting in a pseudogap. For Al, Si, P and S a trio of bands that are degenerate at $\Gamma$ and primarily of $p$-character cross $E_F$. These bands hybridize with the B and N states. Moving across the periodic table the Bader charges on Si, P and S were computed to be +0.42, +0.12 and -0.16, highlighting how the variation of the electronegativity of the enclathrated element affects the direction of charge transfer. The Bader charge on Cl was computed to be -0.44, suggesting that the chemical formula can be viewed as Cl$^-$(B$_5$N$_5$)$^+$, while the noble gas atom in ArB$_5$N$_5$ was nearly neutral, as expected based on its closed electronic shell. The enclathration of Ar within B$_5$N$_5$ increases its PBE band-gap from 4.00~eV to 5.09 for ArB$_5$N$_5$, as expected (\emph{vide supra}).

In addition to Bader charges, we also investigated the Electron Localization Function (ELF) for selected $M$B$_5$N$_5$ phases; typical results are shown in Figure\ \ref{fig:mx10structure}. Unlike the $MX_2$H$_8$ phases derived from $M$H$_{10}$ that form molecular BH$_4^{-}$ and CH$_4$ units, the boron and nitrogen atoms in $M$B$_5$N$_5$ phases maintain a covalent 3-D network similar to the what is observed in compressed $M$H$_{10}$. For instance, in NaB$_5$N$_5$, the ELF between B-N atoms in the clathrate framework reaches a minimum value of 0.85 indicative of a strong covalent bond. Notably this value is higher than between the the H-H atoms in the $Fm\bar{3}m$ LaH$_{10}$ superhydride at 150~GPa (0.64 between the 8$c$ and 32$f$ H atoms and 0.52 between two H atoms at the 32$f$ Wyckoff sites). We also calculated the crystal orbital Hamilton population (COHP) and the negative of the COHP integrated to the Fermi level (-iCOHP) in both AlB$_5$N$_5$ and FB$_5$N$_5$ phases (Table S6). In both structures, the framework is composed of two types of tetrahedral units, BN$_4$ and NB$_4$, which together form the hexagonal faces in the clathrate framework. Notably, the BN$_4$ units exhibit shorter B-N bond distances and larger -iCOHP values, whereas the NB$_4$ units have longer B-N bonds and relatively smaller -iCOHP values. Our -iCOHP analysis also reveals substantial B–N bonding along the square faces of the framework. For instance, in AlB$_5$N$_5$, the –iCOHP values for B–N bonds are 7.16~eV (square face, 1.613~\AA), 7.55~eV (hexagonal face within NB$_4$ units, 1.637~\AA), and 8.84~eV (hexagonal face within BN$_4$ units, 1.557~\AA). Similarly, in FB$_5$N$_5$, the values are 7.64~eV (1.582~\AA), 8.22~eV (1.588~\AA), and 9.37~eV (1.520~\AA) for these same B-N bonds, respectively. These large -iCOHPs support the assignment of a 3-D clathrate-like B–N network in these compounds. In contrast, the intercalated atoms (Al or F) only weakly interact with the surrounding cage atoms, with –iCOHP values of 0.35–0.55~eV for Al–B and 0.1–0.2~eV for Al–N contacts in AlB$_5$N$_5$, and even lower values of 0.10–0.16~eV for F–B and 0.03–0.05~eV for F–N contacts in FB$_5$N$_5$. Visualization of the ELF also supports this interpretation, showing no high ELF values indicative of covalent bonding between the intercalated species and the cage atoms.

The calculated DOS at $E_F$, when the elements in the braces are enclathrated in the boronitride cages, varies between 0.355 (S), 0.130 (Si), 0.129 (Cl), 0.072 (P), 0.054 (Al), 0.044 (Na) states per valence electron, while Mg and Ar filled compounds are predicted to be a semimetal and a wide-gap insulator, respectively. For the metallic systems, typically boron and nitrogen states have similar contributions to the DOS at $E_F$. As the number of $p$ electrons in the intercalant increases, so does its contribution to the metallic bands. The DOS of most boronitride cages filled with transition metal intercalants were not calculated because they were dynamically unstable.  From the explored transition metals those that were stable were found to be semiconductors, as graphically illustrated in Figure \ref{fig:tc}. Ne, Ar, Kr and Xe filled boronitrides had large band gaps of 4.75, 5.09, 5.03 and 4.93~eV, respectively, while the Be-stuffed clathrate cage had an electronic structure similar to that of MgB$_5$N$_5$. 

To analyze the impact of various intercalating elements and the resulting DOS at $E_F$ on the superconducting properties of $M$B$_5$N$_5$ phases, we calculated their $T_c$s. The results are summarized in Figure\ \ref{fig:tc}, where dynamically stable phases are color-coded according to the temperature below which they are estimated to be superconducting. The EPC constant, $\lambda$, and the logarithmic average frequency, $\omega_\text{log}$ are provided in Table S4. 
For this class of compounds, the $\omega_\text{log}$ values ranged between 260~K  to 660~K and most of the $\lambda$ values fell between 0.35 and 0.77, resulting in $T_c$s that  varied from $\sim$1~K to 17~K as calculated via numerical solution of the Eliashberg equations. The computed EPC constants are generally smaller than those previously found for $M$B$_2$C$_8$ phases that are also derived from $M$H$_{10}$~\cite{Li:2024h}, in part because nitrogen is heavier than carbon. Only AlB$_5$N$_5$, FB$_5$N$_5$ and ClB$_5$N$_5$ possessed large $\lambda$ values of 0.88, 1.47 and 0.94, respectively, with predicted $T_c$s of 25~K, 75~K and 46~K, respectively.  
Generally, $\lambda$ decreased and $\omega_\text{log}$ increased moving down the periodic table for each main group element.  The trends in $T_c$ did not fully correspond to the trends in the calculated DOS at $E_F$, which would predict, for example, that SB$_5$N$_5$ should have the highest critical temperature of all of the elements in the third row of the periodic table that were dynamically stable. To understand why this is so, we note that with the exception of the very electropositive elements ($e.g.$ Na and Al) the $p$-states of the enclathrated element contributed significantly to the DOS at $E_F$. However, previous work on MB$_3$C$_3$ and MB$_3$Si$_3$ clathrate systems (and their binary metal analogues)~\cite{Geng:2023, Di:2022, Zhang:2022, Cui:2020} showed that the vibrations of the atoms comprising the cage, and not those of the atoms within the cage ($e.g.$ rattling modes), are responsible for the superconductivity.

\begin{figure*}
\begin{center}
\includegraphics[width=0.8\columnwidth]{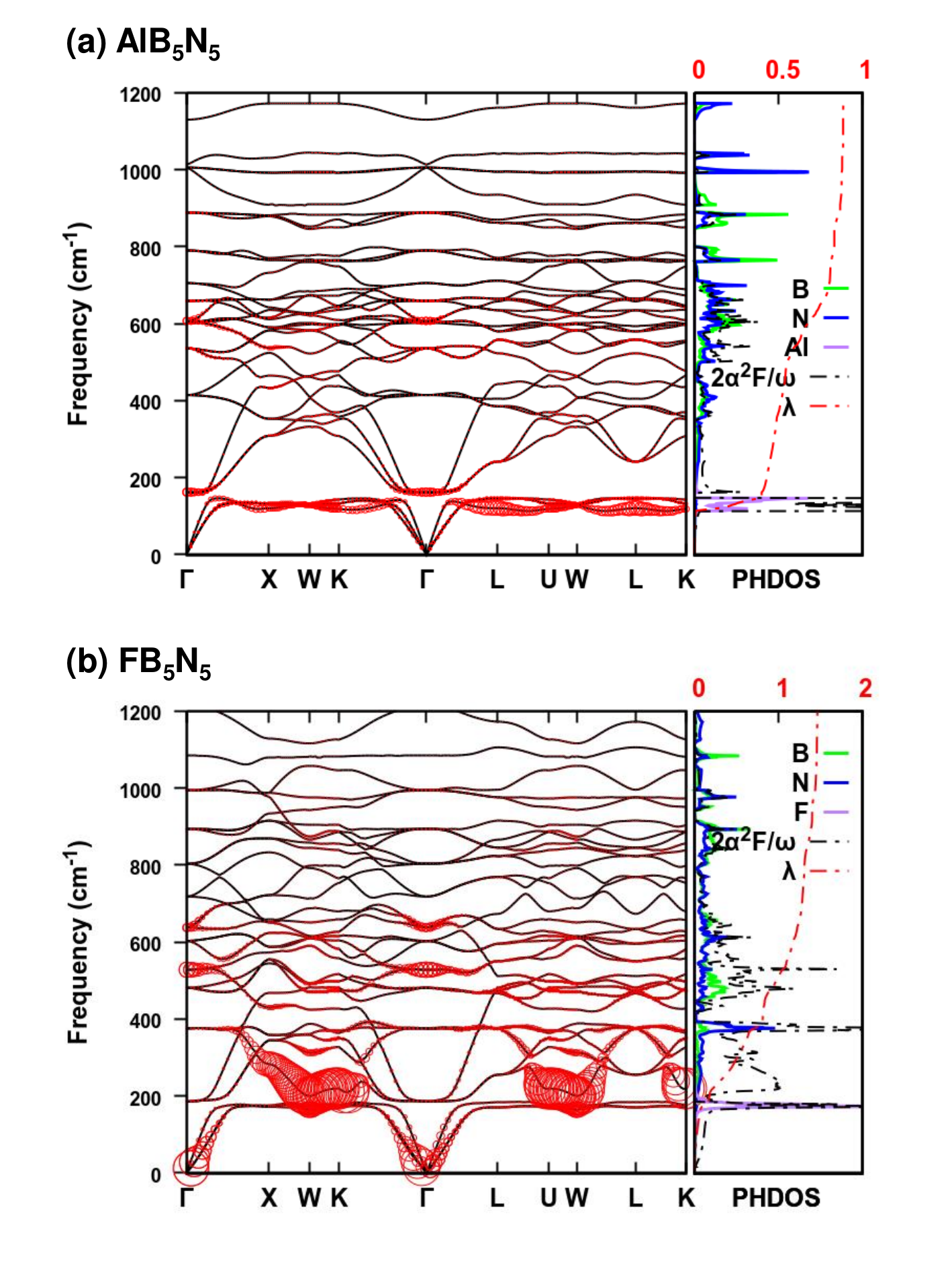}
\caption{Phonon band structures, atom projected phonon density of states (PHDOS), Eliashberg spectral function scaled by the frequency ($2\alpha^2F/\omega$), and the EPC integral ($\lambda(\omega)$) for (a) AlB$_5$N$_5$ and (b) FB$_5$N$_5$. The radius of the bubble on the phonon dispersion curve is proportional to the electron-phonon coupling constant ($\lambda_{\textbf{q}\nu}$) for the mode $\nu$ at wavevector \textbf{q}. }
    \label{fig:epc}
    \end{center}
\end{figure*}

To determine which vibrations contribute most towards the EPC in the  $M$B$_5$N$_5$ compounds, we plotted their phonon band structures with red circles representing the size of the contribution to $\lambda$ at specific wavevectors $\textbf{q}$ and frequencies $\nu$. The results are shown for all of the dynamically stable compounds in Figure S12-14, and for two of the systems with the highest computed $T_c$s, AlB$_5$N$_5$ and FB$_5$N$_5$, in Figure \ref{fig:epc}.  The phonon bands can be divided into a low frequency region (150-250 cm$^{-1}$) associated primarily with vibrations of the intercalated atomic species and the high frequency region (250-1000 cm$^{-1}$) associated primarily with vibrations of the boronitride cages. The contributions to the total $\lambda$ for these regions are provided in Table S4. In general 28-40\% of the contribution to $\lambda$ arises from the three acoustic modes in the low-frequency region for Group 1, 2, and 13-16 elements, while only 10\% of the contribution towards $\lambda$ comes from this region for Group 17 elements. For example, for AlB$_5$N$_5$ (Figure\ \ref{fig:epc}a) these three acoustic branches are essentially dispersionless away from $\Gamma$ with a frequency of around 150-170~cm$^{-1}$, and they are associated with a large EPC. Visualization of phonon modes along these branches within AlB$_5$N$_5$ showed they could be described as rattling motions of the encapsulated Al atoms, and indeed of the enclathrated atoms in the other considered borocarbides. 

As noted above, for ClB$_5$N$_5$ and FB$_5$N$_5$, the findings were slightly different. Since both compounds featured the same soft modes with the largest $\lambda_{\textbf{q}\nu}$ values, only FB$_5$N$_5$ will be discussed here. In FB$_5$N$_5$ the mode with the most significant EPC was associated with soft modes along the  $X$-$W$-$K$ and $U$-$W$ paths, with the $W$ point being particularly notable (Figure\ \ref{fig:epc}b). In the B$_5$N$_5$ cage the boron atoms sit on two different positions, the 16e and the 4c, and those on the 4c (0.25 0.25 0.25) can formally be viewed as the central atoms in the aforementioned BN$_4$ tetrahedral units.  Visualization of the soft modes associated with a large EPC at the  $W$ and $U$ points shows that they corresponded to motions of these BN$_4$ units that resembled the symmetric bending mode in a tetrahedral molecule.

An SrB$_3$C$_3$ phase, analogous to the $M$H$_6$ framework, was recently synthesized under pressure and could be quenched to ambient conditions (though it decomposed upon exposure to air), with a measured $T_c$ of 20~K at 40~GPa and an estimated $T_c$ of 40~K at 1~atm~\cite{zhu:arxiv, Strobel:2021la}. This discovery led us to investigate whether any of the $M$H$_{10}$-like boronitride frameworks predicted in our study could be synthesized.  While assessing dynamic stability is straightforward, it is not the sole criterion for determining the feasibility of new materials with exceptional properties. Kinetic or thermal stability is another crucial aspect, which can be probed via molecular dynamics (MD) simulations. Therefore, we carried out \emph{ab initio} MD simulations at 300~K on the 34 dynamically stable phases, with plots of energy evolution over time provided in Section~S3. Only 11 phases (He, Ne, Ar, Kr, Xe, Pd, Cu, Ag, Zn, Cd, and Hg) remained stable, while the others showed a steep energy decrease within ten picoseconds, indicative of structural transformation. Even these 11 phases, however, exhibited unfavorable formation enthalpies relative to their elemental $M$ phases and BN phases at 0 K from ambient pressure to 100~GPa, not considering potential B/N disorder and the resulting configurational entropy (Figure S15). We also performed MD simulations for FB$_5$N$_5$ at 80 K (slightly higher than the boiling point of liquid nitrogen and the estimated $T_c$ of FB$_5$N$_5$) and we observed that the intercalated F atom prefers to bond with one of the B atoms so that the B-N clathrate framework is twisted and broken (Figure S8). Thus, although we identified several superconducting $M$B$_5$N$_5$ candidates that are dynamically stable at ambient pressure, none of them could maintain kinetic stability at room temperature.

We also predicted the mechanical properties of the dynamically stable $M$B$_5$N$_5$ phases (Table S7), and compared them with that of the optimized empty B$_5$N$_5$ cage. The bulk moduli and shear moduli were estimated using a machine learning model~\cite{Isayev:2017} and calculated with DFT\cite{Wang:2021v}, and the Vickers hardness, $H_\text{v}$, was obtained with the shear modulus coupled with the Teter equation~\cite{Zurek:2019b}. Within this approximation the $H_\text{v}$ of the geometry optimized empty B-N clathrate structure was 27.8 (31.0)~GPa using ML (DFT), which is significantly lower than that of diamond (77.6~GPa) or cubic boron nitride ($c$-BN, 58.4 (57.5)~GPa) calculated~\cite{Zurek:2019b} with the same method. With the exception of FB$_5$N$_5$ the introduction of an atom within the boronitride cage elongates the B-N bonds and increases the volume. This expansion reduces the DFT-Vickers hardness, except in a few cases, notably most of the insulating or semi-conducting $M$B$_5$N$_5$ systems (with the exception of $M=$Rh,  Pd, Pt, Cu), corroborating the expectation that hardness and metallicity are anti-correlated. Inserting the noble gases into the B$_5$N$_5$ cages increases their DFT shear moduli and their Vickers hardness, in-line with previous studies, where computations predicted that insertion of He into small pores found in A-site vacant perovskites increased their bulk moduli~\cite{Zurek:2023b}.

\section{Conclusion}
High-throughput Density Functional Theory (DFT) calculations were used to study 198 $MX_5Y_5$ ($X, Y$ = B, C, or N) clathrate-like structures derived from the $M$H$_{10}$ superhydrides. None of the $M$B$_5$C$_5$ or $M$C$_5$N$_5$ phases were found to be dynamically stable at 1~atm. However, 34 of the $M$B$_5$N$_5$ stoichiometry compounds (Figure 2) were identified as dynamically stable at ambient pressure. With the exception of the noble gases, which remained neutral after intercalation, and the highly electronegative elements, O, S, F, Cl and Br, the intercalating atoms were found to donate electrons to the B-N clathrate framework. FB$_5$N$_5$ was predicted to have the highest ambient pressure superconducting critical temperature, $T_c$, of 75~K. However,  \textit{ab initio} molecular dynamics simulations indicated that all superconducting phases decompose at 300~K at ambient pressure, while only eleven semiconducting phases ($M$ = He, Ne, Ar, Kr, Xe, Pd, Cu, Ag, Zn, Cd and Hg) remained kinetically stable. These findings emphasize the importance of considering kinetic and thermal stability in materials prediction. The often-overlooked aspect of kinetic stability is crucial, and a systematic approach to designing new superconductors through high-throughput methods should be established. DFT calculations generally found the semi-conducting or insulating phases to have a higher hardness than an empty B$_5$N$_5$ cage, with the Vickers hardness of the metallic systems being slightly decreased.

\section*{Declaration of Competing Interest}
The authors declare that they have no known competing financial
interests or personal relationships that could have appeared to influence
the work reported in this paper.

\section*{Data Availability}
The raw/processed data required to reproduce these findings are avalable to download from https://dx.doi.org/10.17172/NOMAD/2025.04.08-2\cite{nomad}. 

\section*{Acknowledgments}
We acknowledge the U.S.\ National Science Foundation for financial support, specifically grants DMR-2119065 (N.G.). Giacomo Scilla acknowledges an Albert Padwa Award from the Department of Chemistry at the University at Buffalo. Calculations were performed at the Center for Computational Research at SUNY Buffalo \cite{ccr}.

\section*{Supporting Information}
Supplementary data to this article can be found online at ... . It includes the computational details, electronic band structures and densities of states, thermodynamic and thermal (molecular dynamics) stability analysis, Bader charges, structural parameters, Eliashberg spectral functions, phonon dispersion curves, EPC calculations, and details of the chemical pressure calculations.



\bibliographystyle{elsarticle-num-names} 
\bibliography{mx5y5}

\begin{thebibliography}{65}
\expandafter\ifx\csname natexlab\endcsname\relax\def\natexlab#1{#1}\fi
\providecommand{\url}[1]{\texttt{#1}}
\providecommand{\href}[2]{#2}
\providecommand{\path}[1]{#1}
\providecommand{\DOIprefix}{doi:}
\providecommand{\ArXivprefix}{arXiv:}
\providecommand{\URLprefix}{URL: }
\providecommand{\Pubmedprefix}{pmid:}
\providecommand{\doi}[1]{\href{http://dx.doi.org/#1}{\path{#1}}}
\providecommand{\Pubmed}[1]{\href{pmid:#1}{\path{#1}}}
\providecommand{\bibinfo}[2]{#2}
\ifx\xfnm\relax \def\xfnm[#1]{\unskip,\space#1}\fi
\bibitem[{Drozdov et~al.(2015)Drozdov, Eremets, Troyan, Ksenofontov, and Shylin}]{Drozdov:2015}
\bibinfo{author}{A.~P. Drozdov}, \bibinfo{author}{M.~I. Eremets}, \bibinfo{author}{I.~A. Troyan}, \bibinfo{author}{V.~Ksenofontov}, \bibinfo{author}{S.~I. Shylin},
\newblock \bibinfo{title}{Conventional superconductivity at 203 {Kelvin} at high pressures in the sulfur hydride system},
\newblock \bibinfo{journal}{Nature} \bibinfo{volume}{525} (\bibinfo{year}{2015}) \bibinfo{pages}{73--76}.
\bibitem[{Ma et~al.(2022)Ma, Wang, Xie, Yang, Wang, Zhou, Liu, Yu, Zhao, Wang et~al.}]{Ma:CaH6}
\bibinfo{author}{L.~Ma}, \bibinfo{author}{K.~Wang}, \bibinfo{author}{Y.~Xie}, \bibinfo{author}{X.~Yang}, \bibinfo{author}{Y.~Wang}, \bibinfo{author}{M.~Zhou}, \bibinfo{author}{H.~Liu}, \bibinfo{author}{X.~Yu}, \bibinfo{author}{Y.~Zhao}, \bibinfo{author}{H.~Wang}, et~al.,
\newblock \bibinfo{title}{High-temperature superconducting phase in clathrate calcium hydride {CaH$_6$} up to 215 {K} at a pressure of 172 {GPa}},
\newblock \bibinfo{journal}{Phys. Rev. Lett.} \bibinfo{volume}{128} (\bibinfo{year}{2022}) \bibinfo{pages}{167001}.
\bibitem[{Li et~al.(2022)Li, He, Zhang, Wang, Zhang, Jia, Feng, Lu, Zhao, Zhang et~al.}]{Li:CaH6}
\bibinfo{author}{Z.~Li}, \bibinfo{author}{X.~He}, \bibinfo{author}{C.~Zhang}, \bibinfo{author}{X.~Wang}, \bibinfo{author}{S.~Zhang}, \bibinfo{author}{Y.~Jia}, \bibinfo{author}{S.~Feng}, \bibinfo{author}{K.~Lu}, \bibinfo{author}{J.~Zhao}, \bibinfo{author}{J.~Zhang}, et~al.,
\newblock \bibinfo{title}{Superconductivity above 200~{K} discovered in superhydrides of calcium},
\newblock \bibinfo{journal}{Nat. Commun.} \bibinfo{volume}{13} (\bibinfo{year}{2022}) \bibinfo{pages}{2863}.
\bibitem[{Semenok et~al.(2021)Semenok, Troyan, Ivanova, Kvashnin, Kruglov, Hanfland, Sadakov, Sobolevskiy, Pervakov, Lyubutin et~al.}]{Semenok:2021}
\bibinfo{author}{D.~V. Semenok}, \bibinfo{author}{I.~A. Troyan}, \bibinfo{author}{A.~G. Ivanova}, \bibinfo{author}{A.~G. Kvashnin}, \bibinfo{author}{I.~A. Kruglov}, \bibinfo{author}{M.~Hanfland}, \bibinfo{author}{A.~V. Sadakov}, \bibinfo{author}{O.~A. Sobolevskiy}, \bibinfo{author}{K.~S. Pervakov}, \bibinfo{author}{I.~S. Lyubutin}, et~al.,
\newblock \bibinfo{title}{Superconductivity at 253~{K} in lanthanum-yttrium ternary hydrides},
\newblock \bibinfo{journal}{Mater. Today} \bibinfo{volume}{48} (\bibinfo{year}{2021}) \bibinfo{pages}{18--28}.
\bibitem[{Kong et~al.(2021)Kong, Minkov, Kuzovnikov, Drozdov, Besedin, Mozaffari, Balicas, Balakirev, Prakapenka, Chariton et~al.}]{Kong:2021a}
\bibinfo{author}{P.~Kong}, \bibinfo{author}{V.~S. Minkov}, \bibinfo{author}{M.~A. Kuzovnikov}, \bibinfo{author}{A.~P. Drozdov}, \bibinfo{author}{S.~P. Besedin}, \bibinfo{author}{S.~Mozaffari}, \bibinfo{author}{L.~Balicas}, \bibinfo{author}{F.~F. Balakirev}, \bibinfo{author}{V.~B. Prakapenka}, \bibinfo{author}{S.~Chariton}, et~al.,
\newblock \bibinfo{title}{Superconductivity up to 243~{K} in the yttrium-hydrogen system under high pressure},
\newblock \bibinfo{journal}{Nat. Commun.} \bibinfo{volume}{12} (\bibinfo{year}{2021}) \bibinfo{pages}{5075}.
\bibitem[{Troyan et~al.(2021)Troyan, Semenok, Kvashnin, Sadakov, Sobolevskiy, Pudalov, Ivanova, Prakapenka, Greenberg, Gavriliuk et~al.}]{Troyan-YH4}
\bibinfo{author}{I.~A. Troyan}, \bibinfo{author}{D.~V. Semenok}, \bibinfo{author}{A.~G. Kvashnin}, \bibinfo{author}{A.~V. Sadakov}, \bibinfo{author}{O.~A. Sobolevskiy}, \bibinfo{author}{V.~M. Pudalov}, \bibinfo{author}{A.~G. Ivanova}, \bibinfo{author}{V.~B. Prakapenka}, \bibinfo{author}{E.~Greenberg}, \bibinfo{author}{A.~G. Gavriliuk}, et~al.,
\newblock \bibinfo{title}{Anomalous high-temperature superconductivity in {YH$_6$}},
\newblock \bibinfo{journal}{Adv. Mater.} \bibinfo{volume}{33} (\bibinfo{year}{2021}) \bibinfo{pages}{2006832}.
\bibitem[{Somayazulu et~al.(2019)Somayazulu, Ahart, Mishra, Geballe, Baldini, Meng, Struzhkin, and Hemley}]{Somayazulu:2019}
\bibinfo{author}{M.~Somayazulu}, \bibinfo{author}{M.~Ahart}, \bibinfo{author}{A.~K. Mishra}, \bibinfo{author}{Z.~M. Geballe}, \bibinfo{author}{M.~Baldini}, \bibinfo{author}{Y.~Meng}, \bibinfo{author}{V.~V. Struzhkin}, \bibinfo{author}{R.~J. Hemley},
\newblock \bibinfo{title}{Evidence for superconductivity above 260~{K} in lanthanum superhydride at megabar pressures},
\newblock \bibinfo{journal}{Phys. Rev. Lett.} \bibinfo{volume}{122} (\bibinfo{year}{2019}) \bibinfo{pages}{027001}.
\bibitem[{Drozdov et~al.(2019)Drozdov, Kong, Minkov, Besedin, Kuzovnikov, Mozaffari, Balicas, Balakirev, Graf, Prakapenka et~al.}]{Drozdov:2019}
\bibinfo{author}{A.~P. Drozdov}, \bibinfo{author}{P.~P. Kong}, \bibinfo{author}{V.~S. Minkov}, \bibinfo{author}{S.~P. Besedin}, \bibinfo{author}{M.~A. Kuzovnikov}, \bibinfo{author}{S.~Mozaffari}, \bibinfo{author}{L.~Balicas}, \bibinfo{author}{F.~F. Balakirev}, \bibinfo{author}{D.~E. Graf}, \bibinfo{author}{V.~B. Prakapenka}, et~al.,
\newblock \bibinfo{title}{Superconductivity at 250~{K} in lanthanum hydride under high pressures},
\newblock \bibinfo{journal}{Nature} \bibinfo{volume}{569} (\bibinfo{year}{2019}) \bibinfo{pages}{528--531}.
\bibitem[{Zurek and Bi(2019)}]{Zurek:2018m}
\bibinfo{author}{E.~Zurek}, \bibinfo{author}{T.~Bi},
\newblock \bibinfo{title}{High-temperature superconductivity in alkaline and rare earth polyhydrides at high pressure: A theoretical perspective},
\newblock \bibinfo{journal}{J. Chem. Phys.} \bibinfo{volume}{150} (\bibinfo{year}{2019}) \bibinfo{pages}{050901}.
\bibitem[{Boeri et~al.(2022)Boeri, Hennig, Hirschfeld, Profeta, Sanna, Zurek, Pickett, Amsler, Dias, Eremets et~al.}]{Zurek:2020k}
\bibinfo{author}{L.~Boeri}, \bibinfo{author}{R.~Hennig}, \bibinfo{author}{P.~Hirschfeld}, \bibinfo{author}{G.~Profeta}, \bibinfo{author}{A.~Sanna}, \bibinfo{author}{E.~Zurek}, \bibinfo{author}{W.~E. Pickett}, \bibinfo{author}{M.~Amsler}, \bibinfo{author}{R.~Dias}, \bibinfo{author}{M.~I. Eremets}, et~al.,
\newblock \bibinfo{title}{The 2021 room-temperature superconductivity roadmap},
\newblock \bibinfo{journal}{J. Phys.: Condens. Matter} \bibinfo{volume}{34} (\bibinfo{year}{2022}) \bibinfo{pages}{183002}.
\bibitem[{Flores-Livas et~al.(2020)Flores-Livas, Boeri, Sanna, Profeta, Arita, and Eremets}]{Flores-Livas:2020}
\bibinfo{author}{J.~A. Flores-Livas}, \bibinfo{author}{L.~Boeri}, \bibinfo{author}{A.~Sanna}, \bibinfo{author}{G.~Profeta}, \bibinfo{author}{R.~Arita}, \bibinfo{author}{M.~Eremets},
\newblock \bibinfo{title}{A perspective on conventional high-temperature superconductors at high pressure: Methods and materials},
\newblock \bibinfo{journal}{Phys. Rep.} \bibinfo{volume}{856} (\bibinfo{year}{2020}) \bibinfo{pages}{1--78}.
\bibitem[{Gao et~al.(2021)Gao, Yan, Lu, and Xiang}]{Gao:2021}
\bibinfo{author}{M.~Gao}, \bibinfo{author}{X.-W. Yan}, \bibinfo{author}{Z.-Y. Lu}, \bibinfo{author}{T.~Xiang},
\newblock \bibinfo{title}{Phonon-mediated high-temperature superconductivity in the ternary borohydride {KB$_2$H$_8$} under pressure near 12 {GPa}},
\newblock \bibinfo{journal}{Phys. Rev. B} \bibinfo{volume}{104} (\bibinfo{year}{2021}) \bibinfo{pages}{L100504}.
\bibitem[{Durajski and Szcz{\c e}{\'s}niak(2021)}]{Durajski:2021}
\bibinfo{author}{A.~P. Durajski}, \bibinfo{author}{R.~Szcz{\c e}{\'s}niak},
\newblock \bibinfo{title}{New superconducting superhydride {LaC$_2$H$_8$} at relatively low stabilization pressure},
\newblock \bibinfo{journal}{Phys. Chem. Chem. Phys.} \bibinfo{volume}{23} (\bibinfo{year}{2021}) \bibinfo{pages}{25070--25074}.
\bibitem[{Hilleke and Zurek(2022)}]{Zurek:2022f}
\bibinfo{author}{K.~Hilleke}, \bibinfo{author}{E.~Zurek},
\newblock \bibinfo{title}{Rational design of superconducting metal hydrides via chemical pressure tuning},
\newblock \bibinfo{journal}{Angew. Chem. Int. Ed.} \bibinfo{volume}{61} (\bibinfo{year}{2022}) \bibinfo{pages}{e202207589}.
\bibitem[{Wan and Zhang(2022)}]{Wan:2022}
\bibinfo{author}{Z.~Wan}, \bibinfo{author}{R.~Zhang},
\newblock \bibinfo{title}{Metallization of hydrogen by intercalating ammonium ions in metal fcc lattices at lower pressure},
\newblock \bibinfo{journal}{Appl. Phys. Lett.} \bibinfo{volume}{121} (\bibinfo{year}{2022}) \bibinfo{pages}{192601}.
\bibitem[{Geng et~al.(2024)Geng, Hilleke, Belli, Das, and Zurek}]{Geng:2024s}
\bibinfo{author}{N.~Geng}, \bibinfo{author}{K.~P. Hilleke}, \bibinfo{author}{F.~Belli}, \bibinfo{author}{P.~K. Das}, \bibinfo{author}{E.~Zurek},
\newblock \bibinfo{title}{Superconductivity in {CH$_4$} and {BH$_4$$^-$} containing compounds derived from the high-pressure superhydrides},
\newblock \bibinfo{journal}{Materials Today Physics} \bibinfo{volume}{44} (\bibinfo{year}{2024}) \bibinfo{pages}{101443}.
\bibitem[{Li et~al.(2022)Li, Wang, Sun, Lu, and Peng}]{Li:2022s}
\bibinfo{author}{S.~Li}, \bibinfo{author}{H.~Wang}, \bibinfo{author}{W.~Sun}, \bibinfo{author}{C.~Lu}, \bibinfo{author}{F.~Peng},
\newblock \bibinfo{title}{Superconductivity in compressed ternary alkaline boron hydrides},
\newblock \bibinfo{journal}{Phys. Rev. B} \bibinfo{volume}{105} (\bibinfo{year}{2022}) \bibinfo{pages}{224107}.
\bibitem[{Jiang et~al.(2022)Jiang, Hai, Tian, Ding, Feng, Yang, Chen, and Zhong}]{Jiang:2022}
\bibinfo{author}{M.-J. Jiang}, \bibinfo{author}{Y.-L. Hai}, \bibinfo{author}{H.-L. Tian}, \bibinfo{author}{H.-B. Ding}, \bibinfo{author}{Y.-J. Feng}, \bibinfo{author}{C.-L. Yang}, \bibinfo{author}{X.-J. Chen}, \bibinfo{author}{G.-H. Zhong},
\newblock \bibinfo{title}{High-temperature superconductivity below 100 {GPa} in ternary {C}-based hydride {MC$_2$H$_8$} with molecular crystal characteristics ({M= Na, K, Mg, Al, and Ga})},
\newblock \bibinfo{journal}{Phys. Rev. B} \bibinfo{volume}{105} (\bibinfo{year}{2022}) \bibinfo{pages}{104511}.
\bibitem[{Zhang et~al.(2022)Zhang, Cui, Hutcheon, Shipley, Song, Du, Kresin, Duan, Pickard, and Yao}]{Zhang:2021}
\bibinfo{author}{Z.~Zhang}, \bibinfo{author}{T.~Cui}, \bibinfo{author}{M.~J. Hutcheon}, \bibinfo{author}{A.~M. Shipley}, \bibinfo{author}{H.~Song}, \bibinfo{author}{M.~Du}, \bibinfo{author}{V.~Z. Kresin}, \bibinfo{author}{D.~Duan}, \bibinfo{author}{C.~J. Pickard}, \bibinfo{author}{Y.~Yao},
\newblock \bibinfo{title}{Design principles for high temperature superconductors with hydrogen-based alloy backbone at moderate pressure},
\newblock \bibinfo{journal}{Phys. Rev. Lett.} \bibinfo{volume}{128} (\bibinfo{year}{2022}) \bibinfo{pages}{047001}.
\bibitem[{Lucrezi et~al.(2022)Lucrezi, Di~Cataldo, von~der Linden, Boeri, and Heil}]{Lucrezi:2022}
\bibinfo{author}{R.~Lucrezi}, \bibinfo{author}{S.~Di~Cataldo}, \bibinfo{author}{W.~von~der Linden}, \bibinfo{author}{L.~Boeri}, \bibinfo{author}{C.~Heil},
\newblock \bibinfo{title}{In-silico synthesis of lowest-pressure high-{$T_c$} ternary superhydrides},
\newblock \bibinfo{journal}{npj Computational Materials} \bibinfo{volume}{8} (\bibinfo{year}{2022}) \bibinfo{pages}{119}.
\bibitem[{Liang et~al.(2021)Liang, Bergara, Wei, Song, Wang, Sun, Liu, Hemley, Wang, Gao, and pthers}]{Liang:2021a}
\bibinfo{author}{X.~Liang}, \bibinfo{author}{A.~Bergara}, \bibinfo{author}{X.~Wei}, \bibinfo{author}{X.~Song}, \bibinfo{author}{L.~Wang}, \bibinfo{author}{R.~Sun}, \bibinfo{author}{H.~Liu}, \bibinfo{author}{R.~J. Hemley}, \bibinfo{author}{L.~Wang}, \bibinfo{author}{G.~Gao}, \bibinfo{author}{pthers},
\newblock \bibinfo{title}{Prediction of high-{$T_c$} superconductivity in ternary lanthanum borohydrides},
\newblock \bibinfo{journal}{Phys. Rev. B.} \bibinfo{volume}{104} (\bibinfo{year}{2021}) \bibinfo{pages}{134501}.
\bibitem[{Di~Cataldo et~al.(2021)Di~Cataldo, Heil, von~der Linden, and Boeri}]{Di:2021bh}
\bibinfo{author}{S.~Di~Cataldo}, \bibinfo{author}{C.~Heil}, \bibinfo{author}{W.~von~der Linden}, \bibinfo{author}{L.~Boeri},
\newblock \bibinfo{title}{{LaBH$_8$}: Towards high-{$T_c$} low-pressure superconductivity in ternary superhydrides},
\newblock \bibinfo{journal}{Phys. Rev. B} \bibinfo{volume}{104} (\bibinfo{year}{2021}) \bibinfo{pages}{L020511}.
\bibitem[{Song et~al.(2023)Song, Bi, Nakamoto, Shimizu, Liu, Zou, Liu, Wang, and Ma}]{Song:2023s}
\bibinfo{author}{Y.~Song}, \bibinfo{author}{J.~Bi}, \bibinfo{author}{Y.~Nakamoto}, \bibinfo{author}{K.~Shimizu}, \bibinfo{author}{H.~Liu}, \bibinfo{author}{B.~Zou}, \bibinfo{author}{G.~Liu}, \bibinfo{author}{H.~Wang}, \bibinfo{author}{Y.~Ma},
\newblock \bibinfo{title}{Stoichiometric ternary superhydride {LaBeH$_8$} as a new template for high-temperature superconductivity at 110 {K} under 80 {GPa}},
\newblock \bibinfo{journal}{Physical Review Letters} \bibinfo{volume}{130} (\bibinfo{year}{2023}) \bibinfo{pages}{266001}.
\bibitem[{Zhu et~al.(2020)Zhu, Borstad, Liu, Gu{\'n}ka, Guerette, Dolyniuk, Meng, Greenberg, Prakapenka, Chaloux et~al.}]{Li:2020sr}
\bibinfo{author}{L.~Zhu}, \bibinfo{author}{G.~M. Borstad}, \bibinfo{author}{H.~Liu}, \bibinfo{author}{P.~A. Gu{\'n}ka}, \bibinfo{author}{M.~Guerette}, \bibinfo{author}{J.-A. Dolyniuk}, \bibinfo{author}{Y.~Meng}, \bibinfo{author}{E.~Greenberg}, \bibinfo{author}{V.~B. Prakapenka}, \bibinfo{author}{B.~L. Chaloux}, et~al.,
\newblock \bibinfo{title}{Carbon-boron clathrates as a new class of $sp^3$-bonded framework materials},
\newblock \bibinfo{journal}{Sci. Adv.} \bibinfo{volume}{6} (\bibinfo{year}{2020}) \bibinfo{pages}{eaay8361}.
\bibitem[{Zhu et~al.(2023)Zhu, Liu, Somayazulu, Meng, Gu{\'n}ka, Shiell, Kenney-Benson, Chariton, Prakapenka, Yoon et~al.}]{zhu:arxiv}
\bibinfo{author}{L.~Zhu}, \bibinfo{author}{H.~Liu}, \bibinfo{author}{M.~Somayazulu}, \bibinfo{author}{Y.~Meng}, \bibinfo{author}{P.~A. Gu{\'n}ka}, \bibinfo{author}{T.~B. Shiell}, \bibinfo{author}{C.~Kenney-Benson}, \bibinfo{author}{S.~Chariton}, \bibinfo{author}{V.~B. Prakapenka}, \bibinfo{author}{H.~Yoon}, et~al.,
\newblock \bibinfo{title}{Superconductivity in srb$_3$c$_3$ clathrate},
\newblock \bibinfo{journal}{Phys. Rev. Res.} \bibinfo{volume}{5} (\bibinfo{year}{2023}) \bibinfo{pages}{013012}.
\bibitem[{Strobel et~al.(2021)Strobel, Zhu, Gu{\'n}ka, Borstad, and Guerette}]{Strobel:2021la}
\bibinfo{author}{T.~A. Strobel}, \bibinfo{author}{L.~Zhu}, \bibinfo{author}{P.~A. Gu{\'n}ka}, \bibinfo{author}{G.~M. Borstad}, \bibinfo{author}{M.~Guerette},
\newblock \bibinfo{title}{A lanthanum-filled carbon-boron clathrate},
\newblock \bibinfo{journal}{Angew. Chem. Int. Ed.} \bibinfo{volume}{60} (\bibinfo{year}{2021}) \bibinfo{pages}{2877--2881}.
\bibitem[{Di~Cataldo et~al.(2022)Di~Cataldo, Qulaghasi, Bachelet, and Boeri}]{Di:2022}
\bibinfo{author}{S.~Di~Cataldo}, \bibinfo{author}{S.~Qulaghasi}, \bibinfo{author}{G.~B. Bachelet}, \bibinfo{author}{L.~Boeri},
\newblock \bibinfo{title}{High-{$T_c$} superconductivity in doped boron-carbon clathrates},
\newblock \bibinfo{journal}{Phys. Rev. B} \bibinfo{volume}{105} (\bibinfo{year}{2022}) \bibinfo{pages}{064516}.
\bibitem[{Zhang et~al.(2022)Zhang, Li, Yang, Wang, Yao, and Liu}]{Zhang:2022}
\bibinfo{author}{P.~Zhang}, \bibinfo{author}{X.~Li}, \bibinfo{author}{X.~Yang}, \bibinfo{author}{H.~Wang}, \bibinfo{author}{Y.~Yao}, \bibinfo{author}{H.~Liu},
\newblock \bibinfo{title}{Path to high-{$T_c$} superconductivity via {Rb} substitution of guest metal atoms in the {SrB$_3$C$_3$} clathrate},
\newblock \bibinfo{journal}{Phys. Rev. B} \bibinfo{volume}{105} (\bibinfo{year}{2022}) \bibinfo{pages}{094503}.
\bibitem[{Gai et~al.(2022)Gai, Guo, Yang, Gao, Gao, and Lu}]{Gai:2022}
\bibinfo{author}{T.-T. Gai}, \bibinfo{author}{P.-J. Guo}, \bibinfo{author}{H.-C. Yang}, \bibinfo{author}{Y.~Gao}, \bibinfo{author}{M.~Gao}, \bibinfo{author}{Z.-Y. Lu},
\newblock \bibinfo{title}{Van hove singularity induced phonon-mediated superconductivity above 77 {K} in hole-doped {SrB$_3$C$_3$}},
\newblock \bibinfo{journal}{Phys. Rev. B} \bibinfo{volume}{105} (\bibinfo{year}{2022}) \bibinfo{pages}{224514}.
\bibitem[{Geng et~al.(2023)Geng, Hilleke, Zhu, Wang, Strobel, and Zurek}]{Geng:2023}
\bibinfo{author}{N.~Geng}, \bibinfo{author}{K.~P. Hilleke}, \bibinfo{author}{L.~Zhu}, \bibinfo{author}{X.~Wang}, \bibinfo{author}{T.~A. Strobel}, \bibinfo{author}{E.~Zurek},
\newblock \bibinfo{title}{Conventional high-temperature superconductivity in metallic, covalently bonded, binary-guest {C-B} clathrates},
\newblock \bibinfo{journal}{J. Am. Chem. Soc.} \bibinfo{volume}{145} (\bibinfo{year}{2023}) \bibinfo{pages}{1696--1706}.
\bibitem[{Li et~al.(2019)Li, Yong, Wu, Lu, Liu, Meng, Tse, and Li}]{Li:2019}
\bibinfo{author}{X.~Li}, \bibinfo{author}{X.~Yong}, \bibinfo{author}{M.~Wu}, \bibinfo{author}{S.~Lu}, \bibinfo{author}{H.~Liu}, \bibinfo{author}{S.~Meng}, \bibinfo{author}{J.~S. Tse}, \bibinfo{author}{Y.~Li},
\newblock \bibinfo{title}{Hard {BN} clathrate superconductors},
\newblock \bibinfo{journal}{J. Phys. Chem. Lett.} \bibinfo{volume}{10} (\bibinfo{year}{2019}) \bibinfo{pages}{2554--2560}.
\bibitem[{Hai et~al.(2022)Hai, Tian, Jiang, Li, Zhong, Yang, Chen, and Lin}]{Hai:2022i}
\bibinfo{author}{Y.-L. Hai}, \bibinfo{author}{H.-L. Tian}, \bibinfo{author}{M.-J. Jiang}, \bibinfo{author}{W.-J. Li}, \bibinfo{author}{G.-H. Zhong}, \bibinfo{author}{C.-L. Yang}, \bibinfo{author}{X.-J. Chen}, \bibinfo{author}{H.-Q. Lin},
\newblock \bibinfo{title}{Improving {$T_c$} in sodalite-like boron-nitrogen compound {M$_2$(BN)$_6$}},
\newblock \bibinfo{journal}{Materials Today Physics} \bibinfo{volume}{25} (\bibinfo{year}{2022}) \bibinfo{pages}{100699}.
\bibitem[{Cui et~al.(2020)Cui, Hilleke, Wang, Lu, Zhang, Zurek, Li, Zhang, Yan, and Bi}]{Cui:2020}
\bibinfo{author}{X.~Cui}, \bibinfo{author}{K.~P. Hilleke}, \bibinfo{author}{X.~Wang}, \bibinfo{author}{M.~Lu}, \bibinfo{author}{M.~Zhang}, \bibinfo{author}{E.~Zurek}, \bibinfo{author}{W.~Li}, \bibinfo{author}{D.~Zhang}, \bibinfo{author}{Y.~Yan}, \bibinfo{author}{T.~Bi},
\newblock \bibinfo{title}{Rb{B}$_{3}${Si}$_{3}$: An alkali metal borosilicide that is metastable and superconducting at 1 atm},
\newblock \bibinfo{journal}{J. Phys. Chem. C} \bibinfo{volume}{124} (\bibinfo{year}{2020}) \bibinfo{pages}{14826--14831}.
\bibitem[{Bi et~al.(2024)Bi, Eggers, Cohen, Campbell, and Strobel}]{Bi:2024}
\bibinfo{author}{T.~Bi}, \bibinfo{author}{B.~T. Eggers}, \bibinfo{author}{R.~E. Cohen}, \bibinfo{author}{B.~J. Campbell}, \bibinfo{author}{T.~Strobel},
\newblock \bibinfo{title}{Computational screening and stabilization of boron-substituted type-{I} and type-{II} carbon clathrates},
\newblock \bibinfo{journal}{J. Am. Chem. Soc.} \bibinfo{volume}{146} (\bibinfo{year}{2024}) \bibinfo{pages}{7985--7997}.
\bibitem[{Li et~al.(2024)Li, Yue, Guo, Zhang, Zhu, Song, Liu, Liu, and Cui}]{Li:2024h}
\bibinfo{author}{J.~Li}, \bibinfo{author}{J.~Yue}, \bibinfo{author}{S.~Guo}, \bibinfo{author}{A.~Zhang}, \bibinfo{author}{L.~Zhu}, \bibinfo{author}{H.~Song}, \bibinfo{author}{Z.~Liu}, \bibinfo{author}{Y.~Liu}, \bibinfo{author}{T.~Cui},
\newblock \bibinfo{title}{High-temperature superconductivity of boron-carbon clathrates at ambient pressure},
\newblock \bibinfo{journal}{Physical Review B} \bibinfo{volume}{109} (\bibinfo{year}{2024}) \bibinfo{pages}{144509}.
\bibitem[{Ding et~al.(2022)Ding, Feng, Jiang, Tian, Zhong, Yang, Chen, and Lin}]{Ding:2022a}
\bibinfo{author}{H.-B. Ding}, \bibinfo{author}{Y.-J. Feng}, \bibinfo{author}{M.-J. Jiang}, \bibinfo{author}{H.-L. Tian}, \bibinfo{author}{G.-H. Zhong}, \bibinfo{author}{C.-L. Yang}, \bibinfo{author}{X.-J. Chen}, \bibinfo{author}{H.-Q. Lin},
\newblock \bibinfo{title}{Ambient-pressure high-${T}_c$ superconductivity in doped boron-nitrogen clathrates {La(BN)$_5$} and {Y(BN)$_5$}},
\newblock \bibinfo{journal}{Physical Review B} \bibinfo{volume}{106} (\bibinfo{year}{2022}) \bibinfo{pages}{104508}.
\bibitem[{Kresse and Hafner(1993)}]{Kresse:1993a}
\bibinfo{author}{G.~Kresse}, \bibinfo{author}{J.~Hafner},
\newblock \bibinfo{title}{\textit{Ab initio} molecular dynamics for liquid metals},
\newblock \bibinfo{journal}{Phys. Rev. B.} \bibinfo{volume}{47} (\bibinfo{year}{1993}) \bibinfo{pages}{558--561}.
\bibitem[{Kresse and Joubert(1999)}]{Kresse:1999a}
\bibinfo{author}{G.~Kresse}, \bibinfo{author}{D.~Joubert},
\newblock \bibinfo{title}{From ultrasoft pseudopotentials to the projector augmented-wave method},
\newblock \bibinfo{journal}{Phys. Rev. B.} \bibinfo{volume}{59} (\bibinfo{year}{1999}) \bibinfo{pages}{1758--1775}.
\bibitem[{Perdew et~al.(1996)Perdew, Burke, and Ernzerhof}]{Perdew:1996a}
\bibinfo{author}{J.~P. Perdew}, \bibinfo{author}{K.~Burke}, \bibinfo{author}{M.~Ernzerhof},
\newblock \bibinfo{title}{Generalized gradient approximation made simple},
\newblock \bibinfo{journal}{Phys. Rev. Lett.} \bibinfo{volume}{77} (\bibinfo{year}{1996}) \bibinfo{pages}{3865--3868}.
\bibitem[{Bl{\"o}chl(1994)}]{Blochl:1994a}
\bibinfo{author}{P.~E. Bl{\"o}chl},
\newblock \bibinfo{title}{Projector augmented-wave method},
\newblock \bibinfo{journal}{Phys. Rev. B} \bibinfo{volume}{50} (\bibinfo{year}{1994}) \bibinfo{pages}{17953--17979}.
\bibitem[{Maintz et~al.(2016)Maintz, Deringer, Tchougr{\'e}eff, and Dronskowski}]{Maintz:2016}
\bibinfo{author}{S.~Maintz}, \bibinfo{author}{V.~L. Deringer}, \bibinfo{author}{A.~L. Tchougr{\'e}eff}, \bibinfo{author}{R.~Dronskowski},
\newblock \bibinfo{title}{Lobster: A tool to extract chemical bonding from plane-wave based {DFT}},
\newblock \bibinfo{journal}{J. Comput. Chem.} \bibinfo{volume}{37} (\bibinfo{year}{2016}) \bibinfo{pages}{1030--1035}.
\bibitem[{Nos{\'e}(1984)}]{Nose:1984}
\bibinfo{author}{S.~Nos{\'e}},
\newblock \bibinfo{title}{A unified formulation of the constant temperature molecular dynamics methods},
\newblock \bibinfo{journal}{J. Chem. Phys.} \bibinfo{volume}{81} (\bibinfo{year}{1984}) \bibinfo{pages}{511--519}.
\bibitem[{Shuichi(1991)}]{Shuichi:1991}
\bibinfo{author}{N.~Shuichi},
\newblock \bibinfo{title}{Constant temperature molecular dynamics methods},
\newblock \bibinfo{journal}{Prog. Theor. Phys. Supp.} \bibinfo{volume}{103} (\bibinfo{year}{1991}) \bibinfo{pages}{1--46}.
\bibitem[{Hoover(1985)}]{Hoover:1985}
\bibinfo{author}{W.~G. Hoover},
\newblock \bibinfo{title}{Canonical dynamics: Equilibrium phase-space distributions},
\newblock \bibinfo{journal}{Phys. Rev. A} \bibinfo{volume}{31} (\bibinfo{year}{1985}) \bibinfo{pages}{1695--1697}.
\bibitem[{Frenkel and Smit(2023)}]{Frenkel:2023}
\bibinfo{author}{D.~Frenkel}, \bibinfo{author}{B.~Smit}, \bibinfo{title}{Understanding molecular simulation: From algorithms to applications}, \bibinfo{publisher}{Elsevier, Amsterdam, Netherlands}, \bibinfo{year}{2023}.
\bibitem[{Parlinski et~al.(1997)Parlinski, Li, and Kawazoe}]{Parlinski:1997}
\bibinfo{author}{K.~Parlinski}, \bibinfo{author}{Z.~Q. Li}, \bibinfo{author}{Y.~Kawazoe},
\newblock \bibinfo{title}{First-principles determination of the soft mode in cubic {ZrO$_2$}},
\newblock \bibinfo{journal}{Phys. Rev. Lett.}  (\bibinfo{year}{1997}) \bibinfo{pages}{4063--4066}.
\bibitem[{Chaput et~al.(2011)Chaput, Togo, Tanaka, and Hug}]{Chaput:2011}
\bibinfo{author}{L.~Chaput}, \bibinfo{author}{A.~Togo}, \bibinfo{author}{I.~Tanaka}, \bibinfo{author}{G.~Hug},
\newblock \bibinfo{title}{Phonon-phonon interactions in transition metals},
\newblock \bibinfo{journal}{Phys. Rev. B} \bibinfo{volume}{84} (\bibinfo{year}{2011}) \bibinfo{pages}{094302}.
\bibitem[{Togo and Tanaka(2015)}]{Togo:2015}
\bibinfo{author}{A.~Togo}, \bibinfo{author}{I.~Tanaka},
\newblock \bibinfo{title}{First principles phonon calculations in materials science},
\newblock \bibinfo{journal}{Scr. Mater.} \bibinfo{volume}{108} (\bibinfo{year}{2015}) \bibinfo{pages}{1--5}.
\bibitem[{Giannozzi et~al.(2009)Giannozzi, Baroni, Bonini, Calandra, Car, Cavazzoni, Ceresoli, Chiarotti, Cococcioni, Dabo et~al.}]{Giannozzi:2009}
\bibinfo{author}{P.~Giannozzi}, \bibinfo{author}{S.~Baroni}, \bibinfo{author}{N.~Bonini}, \bibinfo{author}{M.~Calandra}, \bibinfo{author}{R.~Car}, \bibinfo{author}{C.~Cavazzoni}, \bibinfo{author}{D.~Ceresoli}, \bibinfo{author}{G.~L. Chiarotti}, \bibinfo{author}{M.~Cococcioni}, \bibinfo{author}{I.~Dabo}, et~al.,
\newblock \bibinfo{title}{Quantum espresso: A modular and open-source software project for quantum simulations of materials},
\newblock \bibinfo{journal}{J. Phys.: Condens. Matter} \bibinfo{volume}{21} (\bibinfo{year}{2009}) \bibinfo{pages}{395502}.
\bibitem[{Dal~Corso(2014)}]{DalCorso:2014}
\bibinfo{author}{A.~Dal~Corso},
\newblock \bibinfo{title}{Pseudopotentials periodic table: From {H} to {Pu}},
\newblock \bibinfo{journal}{Comput. Mater. Sci.} \bibinfo{volume}{95} (\bibinfo{year}{2014}) \bibinfo{pages}{337--350}.
\bibitem[{Troullier and Martins(1991)}]{Troullier:1991}
\bibinfo{author}{N.~Troullier}, \bibinfo{author}{J.~L. Martins},
\newblock \bibinfo{title}{Efficient pseudopotentials for plane-wave calculations},
\newblock \bibinfo{journal}{Phys. Rev. B} \bibinfo{volume}{43} (\bibinfo{year}{1991}) \bibinfo{pages}{1993--2006}.
\bibitem[{Monkhorst and Pack(1976)}]{Monkhorst:1976}
\bibinfo{author}{H.~J. Monkhorst}, \bibinfo{author}{J.~D. Pack},
\newblock \bibinfo{title}{Special points for brillouin-zone integrations},
\newblock \bibinfo{journal}{Phys. Rev. B} \bibinfo{volume}{13} (\bibinfo{year}{1976}) \bibinfo{pages}{5188}.
\bibitem[{Methfessel and Paxton(1989)}]{Methfessel:1989}
\bibinfo{author}{M.~Methfessel}, \bibinfo{author}{A.~T. Paxton},
\newblock \bibinfo{title}{High-precision sampling for brillouin-zone integration in metals},
\newblock \bibinfo{journal}{Phys. Rev. B} \bibinfo{volume}{40} (\bibinfo{year}{1989}) \bibinfo{pages}{3616--3621}.
\bibitem[{Allen and Dynes(1975)}]{Allen:1975}
\bibinfo{author}{P.~B. Allen}, \bibinfo{author}{R.~C. Dynes},
\newblock \bibinfo{title}{Transition temperature of strong-coupled superconductors reanalyzed},
\newblock \bibinfo{journal}{Phys. Rev. B} \bibinfo{volume}{12} (\bibinfo{year}{1975}) \bibinfo{pages}{905--922}.
\bibitem[{Eliashberg(1960)}]{Eliashberg:1960}
\bibinfo{author}{G.~M. Eliashberg},
\newblock \bibinfo{title}{Interactions between electrons and lattice vibrations in a superconductor},
\newblock \bibinfo{journal}{Sov. Phys. JETP} \bibinfo{volume}{11} (\bibinfo{year}{1960}) \bibinfo{pages}{696--702}.
\bibitem[{Zhang et~al.(2013)Zhang, Wang, Lv, Zhu, Li, Zhang, Li, and Ma}]{Zhang:2013a}
\bibinfo{author}{X.~Zhang}, \bibinfo{author}{Y.~Wang}, \bibinfo{author}{J.~Lv}, \bibinfo{author}{C.~Zhu}, \bibinfo{author}{Q.~Li}, \bibinfo{author}{M.~Zhang}, \bibinfo{author}{Q.~Li}, \bibinfo{author}{Y.~Ma},
\newblock \bibinfo{title}{First-principles structural design of superhard materials},
\newblock \bibinfo{journal}{J. Chem. Phys.} \bibinfo{volume}{138} (\bibinfo{year}{2013}) \bibinfo{pages}{114101}.
\bibitem[{Racioppi et~al.(2023)Racioppi, Miao, and Zurek}]{Zurek:2023b}
\bibinfo{author}{S.~Racioppi}, \bibinfo{author}{M.~Miao}, \bibinfo{author}{E.~Zurek},
\newblock \bibinfo{title}{Intercalating helium into {A}-site vacant perovskites},
\newblock \bibinfo{journal}{Chem. Mater.} \bibinfo{volume}{35} (\bibinfo{year}{2023}) \bibinfo{pages}{4297--4310}.
\bibitem[{Hester et~al.(2017)Hester, {Dos Santos}, Molaison, Hancock, and Wilkinson}]{Hester_2017_JACS}
\bibinfo{author}{B.~R. Hester}, \bibinfo{author}{A.~M. {Dos Santos}}, \bibinfo{author}{J.~J. Molaison}, \bibinfo{author}{J.~C. Hancock}, \bibinfo{author}{A.~P. Wilkinson},
\newblock \bibinfo{title}{{Synthesis of defect perovskites (He$_{2-x}$$\square$$_x$)(CaZr)F$_6$ by inserting helium into the negative thermal expansion material CaZrF$_6$}},
\newblock \bibinfo{journal}{J. Am. Chem. Soc.} \bibinfo{volume}{139} (\bibinfo{year}{2017}) \bibinfo{pages}{13284--13287}.
\bibitem[{Lloyd et~al.(2021)Lloyd, Hester, Baxter, Ma, Prakapenka, Tkachev, Park, and Wilkinson}]{Lloyd_2021_ChemofMat}
\bibinfo{author}{A.~J. Lloyd}, \bibinfo{author}{B.~R. Hester}, \bibinfo{author}{S.~J. Baxter}, \bibinfo{author}{S.~Ma}, \bibinfo{author}{V.~B. Prakapenka}, \bibinfo{author}{S.~N. Tkachev}, \bibinfo{author}{C.~Park}, \bibinfo{author}{A.~P. Wilkinson},
\newblock \bibinfo{title}{{Hybrid double perovskite containing helium: [He$_2$][CaZr]F$_6$}},
\newblock \bibinfo{journal}{Chem. Mater.} \bibinfo{volume}{33} (\bibinfo{year}{2021}) \bibinfo{pages}{3132--3138}.
\bibitem[{Shipley et~al.(2021)Shipley, Hutcheon, Needs, and Pickard}]{Shipley:2021a}
\bibinfo{author}{A.~M. Shipley}, \bibinfo{author}{M.~J. Hutcheon}, \bibinfo{author}{R.~J. Needs}, \bibinfo{author}{C.~J. Pickard},
\newblock \bibinfo{title}{High-throughput discovery of high-temperature conventional superconductors},
\newblock \bibinfo{journal}{Phys. Rev. B} \bibinfo{volume}{104} (\bibinfo{year}{2021}) \bibinfo{pages}{054501}.
\bibitem[{Isayev et~al.(2017)Isayev, Oses, Toher, Gossett, Curtarolo, and Tropsha}]{Isayev:2017}
\bibinfo{author}{O.~Isayev}, \bibinfo{author}{C.~Oses}, \bibinfo{author}{C.~Toher}, \bibinfo{author}{E.~Gossett}, \bibinfo{author}{S.~Curtarolo}, \bibinfo{author}{A.~Tropsha},
\newblock \bibinfo{title}{Universal fragment descriptors for predicting properties of inorganic crystals},
\newblock \bibinfo{journal}{Nat. Commun.} \bibinfo{volume}{8} (\bibinfo{year}{2017}) \bibinfo{pages}{15679}.
\bibitem[{Wang et~al.(2021)Wang, Xu, Liu, Tang, and Geng}]{Wang:2021v}
\bibinfo{author}{V.~Wang}, \bibinfo{author}{N.~Xu}, \bibinfo{author}{J.-C. Liu}, \bibinfo{author}{G.~Tang}, \bibinfo{author}{W.-T. Geng},
\newblock \bibinfo{title}{{VASPKIT}: {A} user-friendly interface facilitating high-throughput computing and analysis using {VASP} code},
\newblock \bibinfo{journal}{Comput. Phys. Commun.} \bibinfo{volume}{267} (\bibinfo{year}{2021}) \bibinfo{pages}{108033}.
\bibitem[{Avery et~al.(2019)Avery, Wang, Oses, Gossett, Proserpio, Toher, Curtarolo, and Zurek}]{Zurek:2019b}
\bibinfo{author}{P.~Avery}, \bibinfo{author}{X.~Wang}, \bibinfo{author}{C.~Oses}, \bibinfo{author}{E.~Gossett}, \bibinfo{author}{D.~Proserpio}, \bibinfo{author}{C.~Toher}, \bibinfo{author}{S.~Curtarolo}, \bibinfo{author}{E.~Zurek},
\newblock \bibinfo{title}{Predicting superhard materials via a machine learning informed evolutionary structure search},
\newblock \bibinfo{journal}{npj Comput. Mater.} \bibinfo{volume}{5} (\bibinfo{year}{2019}) \bibinfo{pages}{89}.
\bibitem[{\emph{Novel Materials Discovery.}(2025)}]{nomad}
\bibinfo{author}{\emph{Novel Materials Discovery.}}, \bibinfo{year}{2025}. \bibinfo{note}{Https://nomad-lab.eu/, DOI:https://dx.doi.org/10.17172/NOMAD/2025.04.08-2 (accessed April 8th, 2025)}.
\bibitem[{\text{Center for Computational Research, University at Buffalo.}(2025)}]{ccr}
\bibinfo{author}{\text{Center for Computational Research, University at Buffalo.}}, \bibinfo{year}{2025}. \bibinfo{note}{Http://hdl.handle.net/10477/79221 (accessed April 8th, 2025)}.

\end{thebibliography}


\end{document}